\documentclass{article}
\usepackage{spconf}
\usepackage{cite}
\usepackage{amsmath,amssymb,amsfonts}
\usepackage{algorithm}
\usepackage{algpseudocode}\usepackage{graphicx}
\usepackage{textcomp}
\usepackage{xcolor}
\usepackage{bm}
\usepackage{subcaption}
\usepackage{makecell}
\usepackage{comment}
\usepackage[export]{adjustbox}

\name{Christian D. Rask, Daniel E. Lucani\thanks{This work was supported in part by the GROWLean Project under Grant DFF-2035-00229B granted by the Independent Research Fund Denmark. The authors gratefully acknowledge STMicroelectronics for their constructive collaboration. Special thanks to the Embedded Graphics team for their expertise and support, which contributed to the findings in this research.}}
\address{Department of Electrical and Computer Engineering, Aarhus University, Denmark \\ email: christianrask2@gmail.com, daniel.lucani@ece.au.dk}

\begin{document}

\title{RAGE for the Machine: Image Compression with Low-Cost Random Access for Embedded Applications}


\maketitle

\begin{abstract}
We introduce RAGE, an image compression framework that achieves four generally conflicting objectives: 1) good compression for a wide variety of color images, 2) computationally efficient, fast decompression, 3) fast random access of images with pixel-level granularity without the need to decompress the entire image, 4) support for both lossless and lossy compression. To achieve these, we rely on the recent concept of generalized deduplication (GD), which is known to provide efficient lossless (de)compression and fast random access in time-series data, and deliver key expansions suitable for image compression, both lossless and lossy. Using nine different datasets, incl. graphics, logos, natural images, we show that RAGE has similar or better compression ratios to state-of-the-art lossless image compressors, while delivering pixel-level random access capabilities. Tests in an ARM Cortex-M33 platform show seek times between 9.9 and 40.6~ns and average decoding time per pixel between 274 and 1226~ns. Our measurements also show that RAGE's lossy variant, RAGE-Q, outperforms JPEG by several fold in terms of distortion in embedded graphics and has reasonable compression and distortion for natural images. 
\end{abstract}

\section{Introduction}
\begin{figure}
    \centering
    \includegraphics[width=0.37\textwidth]{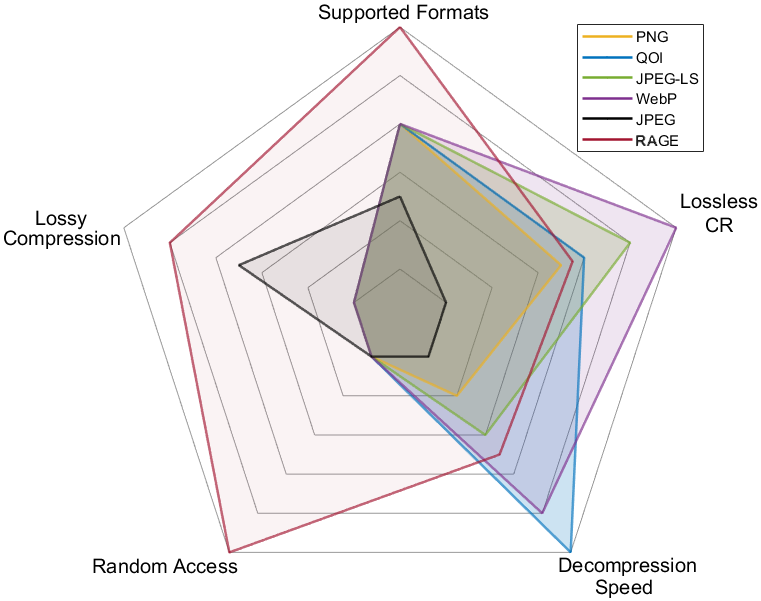}
    \caption{RAGE performance benefits in different dimensions}
    \label{fig:RAGE-benefits}
\end{figure}

In recent years, the demand for Graphical User Interfaces (GUI) on resource-constrained embedded devices has witnessed significant growth due to the high demand for user-friendly and visually appealing interfaces \cite{gui}. When rendering GUIs, only a subset of each frame must be rendered due to high temporal correlation between frames. GUI software development tools for embedded devices exploit this fact by computing the areas eligible for re-drawing between each frame, and only render these for efficiency. This spawns the need for low-cost random access to images used in the GUI. Traditionally, embedded graphics applications have stored image data in a raw format to enable random access. However, storing uncompressed images can quickly consume the limited read-only memory found on such devices. One approach to reduce memory consumption is to store the images in a compressed format. However, most traditional image compressors do not suffice due to their poor support for random access queries, implying much of the compressed data must be decompressed to retrieve even a single pixel. An image compression scheme that enables access to the compressed data without complete decompression is thus desired.

The overall goals of an embedded graphics image compressor are to \textbf{(a)} achieve good compression of color images, \textbf{(b)} be computationally efficient and allow for fast decompression on the target side, and \textbf{(c)} allow for low-cost random access to the compressed data, enabling efficient queries and retrieval.
These goals conflict, e.g., facilitating random access might compromise the achieved compression ratios. Thus, a scheme that achieves a good tradeoff between them and that can be deployed in a constrained, embedded device is desired.

\textbf{Fine-grained Random Access:} 
\begin{figure}
    \centering
    \includegraphics[width=0.39\textwidth]{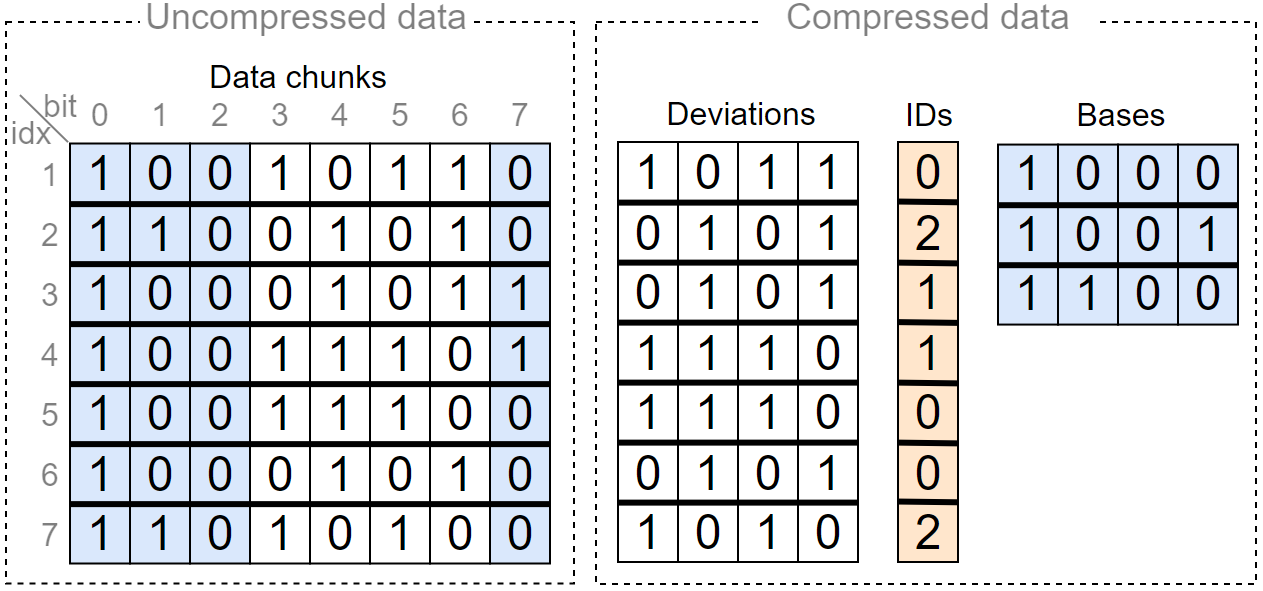}
    \caption{GD applied to 8-bit data chunks. Bits in blue are allocated to the bases and deduplicated. The remaining bits are deviations stored alongside an associated base ID.}
    \label{fig:gd_example}
\end{figure}
Recently, generalized deduplication (GD) \cite{9014012} has been shown to achieve the aforementioned goals, particularly, on time-series data in \cite{9155450, 9435794}. These show that the compression ratios are comparable to several universal compressors, while being fast, providing low-cost random access capabilities, and being practical for simple real-time IoT devices. GD compresses data by transforming fixed-sized \textit{chunks}, $\bm{c}$, into \textit{bases}, $\bm{b}$, which are deduplicated, and \textit{deviations}, $\bm{d}$. Deviations are stored alongside a \textit{base ID} that links them to their associated base (Fig.~\ref{fig:gd_example}).
This allows for lossless decompression of the data and low-cost random access, since base ID and deviation pairs have a fixed size. Compressing data with GD has two stages: 1) configuration, where parameters are determined, and 2) compression, where each chunk is transformed, and bases are deduplicated.

\begin{figure*}
    \centering
    \includegraphics[width=\textwidth]{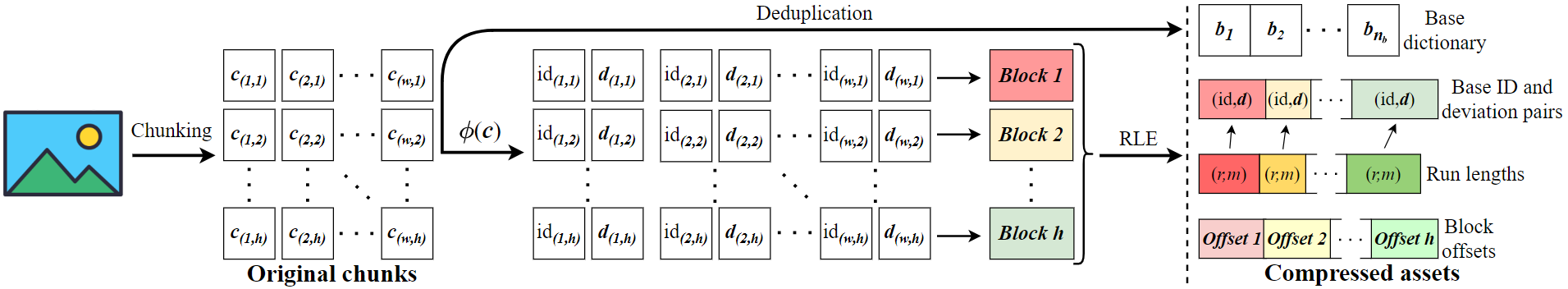}
    \caption{RAGE data flow. An image, $I$, is split into $n$ chunks. Chunks are transformed and bases are deduplicated. Each row of $(\text{id},\bm{d})$-pairs are collected into $h$ blocks. RLE is applied to each block independently. The compressed image consists of several compressed assets needed to recreate the original image.}
    \label{fig:data_flow}
    \vspace{-12pt}
\end{figure*}

\textbf{Contribution:} In this work we propose the \textit{Randomly Accessible Graphics Encoding} (\textbf{RAGE}) to losslessly compress color images using GD. To the best of our knowledge, this is the first application of GD to image compression. 
We combine GD with a variation of Run Length Encoding (RLE) to more efficiently compress \textit{discrete-tone} images, while still enabling efficient random access querying. We also propose a lossy variant, which we call RAGE-Q (Quantized), that allows for higher compression with acceptable visual distortion for discrete-tone images. RAGE benefits over other compressors are summarized in Fig.~\ref{fig:RAGE-benefits}.

\section{Related Work}

Locally decodable compression has attracted significant attention from a theoretical perspective~\cite{5394919, 8437931, 9840790}. For example, \cite{8437931} proposes a generic scheme that adapts any universal compressor into one with constant-time random access. These proposals focus on the fundamental principles and asymptotic limits of compressors with random access capabilities, but their structure makes them unsuitable for real applications.
The simplest (and most common) approach to adapt a compressor to have random access is to split the source data stream into multiple data blocks and compress each data block independently. Recent works have studied trade-offs between block size, compression ratio, and random access speed of block-based methods, e.g.,~\cite{9840790}, and developed practical frameworks to adapt any universal compressor into one with improved random access capabilities~\cite{9484460}. 
Thus, random access granularity is given by the block size. The goal is to reduce the block size while maintaining good compression.

Image compression algorithms typically deploy a chunk-based approach to encode images, e.g., one-dimensional row-based blocks in lossless PNG \cite{png}, 
small two-dimensional pixel blocks in JPEG \cite{jpeg}.
However, image compressors are generally designed to minimize image size and do not support random access.
Recently, GD-based data compression has been proposed to support low-cost random access with good compression ratios. For multidimensional data, GreedyGD~\cite{greedygd} provides a systematic, greedy algorithm to select parameters balancing good compression, random access and support accurate analytics to be performed directly on the compressed data. GreedyGD uses an efficient tree structure (BaseTree) to represent bases during the configuration stage and select optimal parameters for the input data.

\section{Randomly Accessible Graphics Encoding}

\subsection{System Model}
We consider an image, $I$, to contain $n$ chunks, each of $l_c$ bits
Each chunk will be processed by a transformation function, $\phi(\cdot)$, that splits each chunk into a base and a deviation.
We denote \textit{base bits} ($B$) as the set of bit positions that are allocated to bases, while the remaining are \textit{deviation bits}. 
The number of bits of a base is then $l_b = |B|$, while each deviation has a total  $l_d=l_c-l_b$ bits.
This transformation is bijective.
We use vector notation to represent chunks of binary data denoted with small boldface letters, i.e., $\bm{v}$, and the number of binary values as $l_v$. 

Prior to compressing, the input image, $I$, is passed to a configuration stage where we select the bit positions to allocate to bases, $B$, and the length of bases, $l_b$. An optional quantization step can be performed for lossy compression (see section \ref{sec:lossy_compression}). Next, the compression starts. The overall data flow of the compression pipeline is described in Fig. \ref{fig:data_flow}.

The \textit{pixel} defines the granularity of the chunks to be processed. Therefore, the size of each chunk, $l_c$, of an image is $l_c=\text{bpp}$, which is the bits per pixel (bpp) of the image (typically 24 or 32). An image with dimensions $w \times h$ has $n=w\cdot h$ chunks and each can be indexed as either an index to a 1-D raster-scan array of all pixels, $\bm{c}_i$, or as a 2-D coordinate, $\bm{c}_{(x,y)}$. Chunks are passed through the transformation function, $\phi(\bm{c}_i)$, which split each chunk into a base, $\bm{b}_i$ and deviation, $\bm{d}_i$. All bases are deduplicated and all unique instances are stored in a base dictionary, $\mathcal{M}$, and are each assigned a unique base ID. The number of bits used per base ID is $l_{id}=\lceil \log_2{n_b} \rceil$.

Next, all $(\text{id},\bm{d})$-pairs of individual \textit{rows} of the image are group together in blocks. Therefore, we have $h$ blocks in total. RLE is applied to each block independently. We record the start offsets of each encoded $(\text{id},\bm{d})$ block for locating independent compressed blocks. The compressed image, $I_C$, consists of several assets needed to reconstruct the original image. We have left out the parameters needed by the decoder, since the size of these is typically negligible. 
By applying RLE to the $(\text{id},\bm{d})$-pairs we compromise the cost of random access queries which GD supports. However, by encoding blocks we have access to individual rows of pairs. To get access to a given chunk, the decoder must (1) read block offsets, (2) accumulate RLE values, and (3) fetch the associated $(\text{id},\bm{d})$ pair and decode this.

\subsection{Base Selection Algorithm}

\begin{figure}
    \centering
    \includegraphics[width=0.48\textwidth]{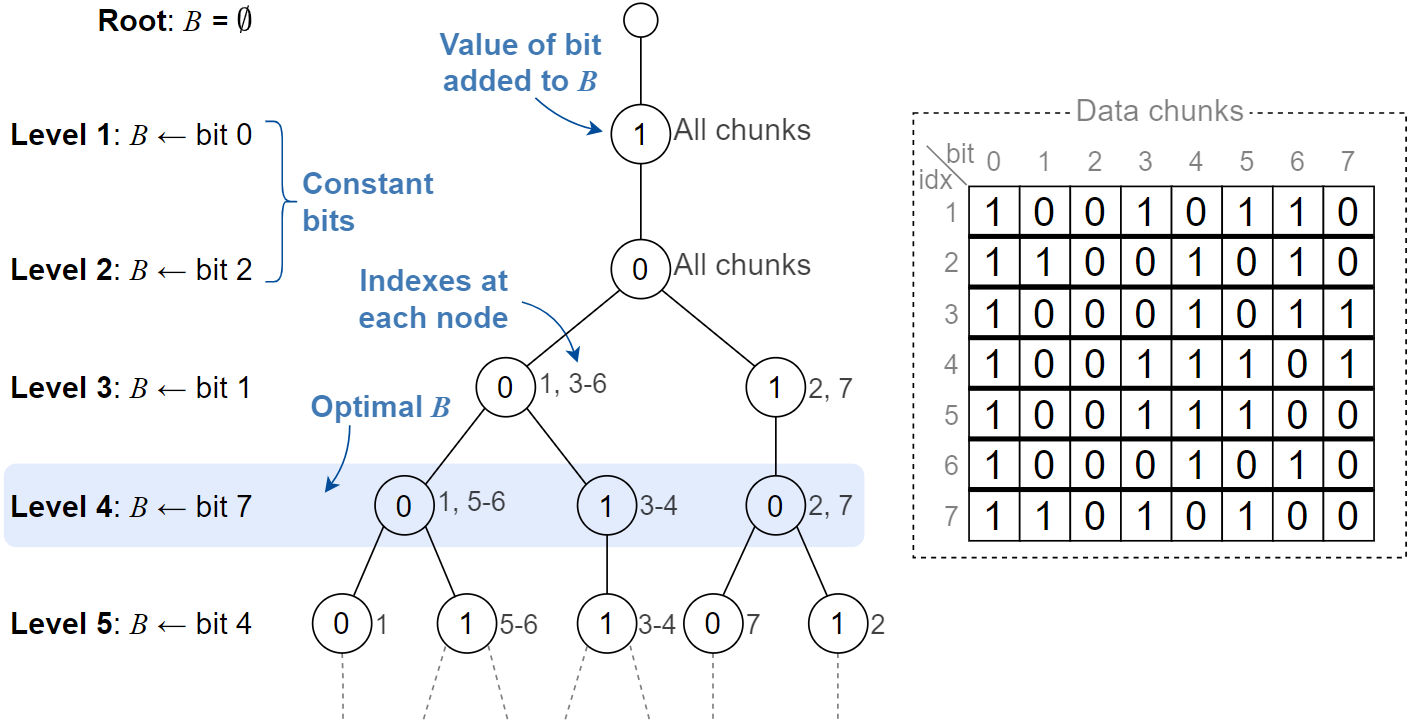}
    \caption{Modified \texttt{BaseTree} applied to the data chunks in Fig. \ref{fig:gd_example}. Level 4 is highlighted as the optimal set of base bits $B$, which corresponds to the base and deviation bits in Fig.~\ref{fig:gd_example}.}
    \label{fig:basetree}
    \vspace{-12pt}
\end{figure}

To enable efficient selection of high quality base bits in the configurations stage, the data is organized according to the \texttt{BaseTree} structure described in \cite{greedygd} (see Fig.~\ref{fig:basetree}).
The width of the tree at any level corresponds to $n_b$ for a given set of base bits $B$. Furthermore, the height of the tree corresponds to $l_b$ and each root-to-leaf path to individual bases. The leaf nodes at any level record which data chunk indexes are mapped to each base.

To find an optimal base length, $l_b$, the selection algorithm described in Alg. \ref{alg:selection} is used.
\begin{figure}[t]
\vspace{-24pt}
\begin{algorithm}[H]
\small
\begin{algorithmic}[1]
\Require Image $I$, chunk size $l_c$
\Ensure Base bits $B$
\State $B \gets \text{constant bits in } I$ \Comment{Initialization}
\State Expand \texttt{BaseTree} with $B$
\State $B_{best} \gets B$
\State $S_{best} \gets \infty$
\While{$|B| < l_c$} \Comment{Iteratively add base bits}
    \State $S_{loc} \gets \infty$
    \State $b_i \gets$ bit position spawning least new bases.
    \State Expand \texttt{BaseTree} with $b_i$
    \State $B \gets B \cup \{ b_i \}$
    \State $S_{loc} \gets$ compressed size \Comment{Eq. \eqref{eq:gd_size_total}}
    \If{$S_{loc} < S_{best}$} \Comment{Update global best}
        \State $B_{best} \gets B$
        \State $S_{best} \gets S_{loc}$
    \EndIf
\EndWhile 
\State \Return $B_{best}$
\end{algorithmic}
\caption{Base bit selection algorithm} \label{alg:selection}
\end{algorithm}
\vspace{-26pt}
\end{figure}
The process begins by adding all constants bits if any to $B$ and expanding \texttt{BaseTree}.
In line 7 we select the next bit, $b_i \notin B$, which when added to \texttt{BaseTree} results in the smallest $n_b$. The chosen bit is added to $B$ and we expand \texttt{BaseTree} accordingly. The local estimated compressed size, $\S_{loc}$, is calculated and compared against the global optimum. It becomes the new global optimum if smaller. When all bit positions have been added to $B$, the process terminates and return the state of the best configuration. By iteratively choosing greedy base bits and estimating the compressed size, we effectively achieve high compression rates, while keeping the number of bases low and the deviations small.

\subsection{Run Length Encoding (RLE)}

Graphical (discrete-tone) images often used in GUIs are characterized by uniform regions separated by sharp contrasts as opposed to continuous-tone natural photos. Thus, RLE is a natural choice for compressing graphical images. RLE generally provides worse compression performance compared to its more generalized forms, e.g., those based on LZ77 \cite{1055714}. However, decoding a compressed back-referencing stream entails that the decoder must maintain all prior symbols in the search window to interpret the next. This compromises the random-access speed. For RLE data, only the run lengths are needed, and the decoder accumulates run lengths until it reaches the start of a query area while ignoring the associated symbols which are not needed.

We propose an RLE scheme that efficiently encodes symbol runs but also sequences without runs. The RLE sequence is initialized with a run length, $r_1$. Its associated symbol is stored in a separate stream. Next follows a number, $m_1$, indicating that the next group of $m_1$ symbols are all different (i.e., no runs). The group of $m_1$ different symbols is appended to the symbol stream. After this group, we encode a potential run again. The run length $r_2$ and its symbol are added to the stream. Next follows another group of $m_2$ different values \cite{handbook}. This alternating pattern of $r$- and $m$-values continues until the end of the data block. 
$r$-values are encoded with a bias of -1 as we do not encode runs under 2 as run lengths.

\subsection{Lossy Encoding via Tree Pruning  }\label{sec:lossy_compression}

We now describe the optional quantization step, which is performed as part of the configuration stage of the lossy RAGE-Q scheme. It uses the \texttt{BaseTree} structure to identify less relevant bases that account for only a small amount of the total information of a given source image and prune these. We refer to this process as Base Tree Pruning (BTP). Pruning a base result in its associated chunks being mapped to another base by flipping specific bits of the chunks. By pruning less important bases we have a smaller base dictionary, $\mathcal{M}$, and potentially decrease $l_{id}$.

\begin{figure}
    \centering
    \includegraphics[width=0.48\textwidth]{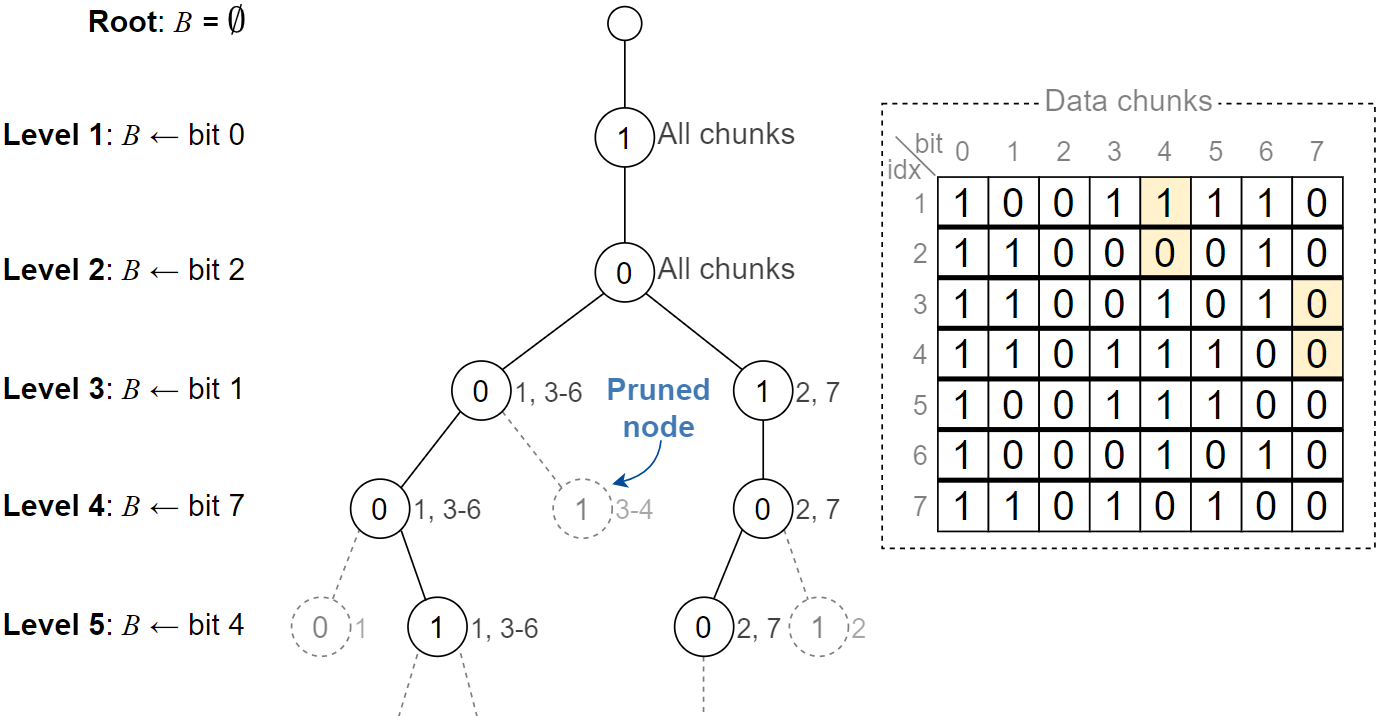}
    \caption{BTP applied to \texttt{BaseTree} from Fig. \ref{fig:basetree}. Flipped bits are marked yellow. $\text{PSNR}_{thr} = 30$.}
    \label{fig:basetree_pruning}
\end{figure}

When constructing \texttt{BaseTree}, each parent node either spawns one or two children nodes. When two children are spawned, the width of the tree expands, i.e., we increase $n_b$. A weighted mapping direction based on the number of chunks that are mapped to each child is defined. The child with the most chunks mapped to it becomes the target node which the other child is mapped to if pruned. To decide whether to prune, we apply a cost function that quantifies the amount of distortion introduced by performing the mapping. We define the \textit{binary mean square error} (BMSE) as
\begin{equation}
    \vspace{-4pt}
    \text{BMSE} \triangleq \frac{1}{n} \sum_{i=1}^{k}{(2^mp_i - 2^mq_i)^2},
    \vspace{-2pt}
\end{equation}
where $n$ is the total number of chunks in the image, $k$ is the number of chunks mapped the node to prune, $m$ is the bit significance (relative to the channel) of the bit position of the current level, and $p$ and $q$ are binary values. We define the \textit{peak signal to noise ratio} (PSNR) as
\begin{equation}
    \text{PSNR} \triangleq 20 \log_{10}{\frac{2^t}{\text{RBMSE}}},
\end{equation}
where RBMSE is the root BMSE and $t$ denotes the highest significance a bit in a channel can have (e.g., $t=7$ for 8-bit channels). The PSNR cost is calculated for each child node to be pruned in each subsequent level of \texttt{BaseTree}. Users of the algorithm specify a quality parameter which is a threshold value, $\text{PSNR}_{thr}$. If the PSNR cost of a given base mapping is greater than $\text{PSNR}_{thr}$ the mapping is performed. Fig. \ref{fig:basetree_pruning} shows BTP applied to the base tree in Fig. \ref{fig:basetree}.

\subsection{Compressed Size}

The size of the compressed data is now analyzed. The $r$- and $m$ values for the RLE scheme compose an ordered sequence in an alternating pattern as ${\cal A} \triangleq (r_1, m_1, r_2, m_2, \dots) = (a_0, ..., a_{|{\cal A}|-1}).$
Each value is encoded as 4- or 8-bit packets to have a compact representation and enable fast unpacking. The first bit of each packet indicates whether the next 3 or 7 bits contain the value. Values in the range $[0,8)$ are encoded as 4-bit packets. Larger values are encoded as 8-bit packets. A bias of $-8$ is applied, meaning we can store values in the range $[8,136)$ using 7 bits in 8-bit packets. The size in bits of any $r$- or $m$-value is
\begin{equation}
RLE(v) \triangleq
    \begin{cases}
        4 & \text{, if } 0 \leq v < 8 \\
        8 & \text{, if } 8 \leq v < 136 
    \end{cases},
\end{equation}
where $v$ is any $r$- or $m$ value. We require that the encoder splits values larger than $136$ into smaller $r$- or $m$-values. The size in bits of ${\cal A}$ is
\begin{equation}
    S_{RLE} \triangleq \sum_{k=0}^{|{\cal A}|-1}RLE(a_k).
\end{equation}
For each run length $r_i \neq 0$ one $(\text{id},\bm{d})$-pair is present in the pair stream. For each difference value, $m_i$, there are $m_i$ $(\text{id},\bm{d})$-pairs. The number of $(\text{id},\bm{d})$-pairs is
\begin{equation}
    N_{pairs} \triangleq \sum_{k=0}^{|{\cal A}| / 2 - 1}{\left( \bm{1}[a_{2k} \neq 0] + a_{2k+1} \right)},
\end{equation}
where $\bm{1}[\cdot]$ is an indicator function, taking a value of 1 if the argument is true, and 0 otherwise. The size of storing all $(\text{id},\bm{d})$-pairs is
\begin{equation}\label{eq:pair_size}
    S_{pairs} \triangleq N_{pairs} ( l_{id} + l_d ).
\end{equation}
We must also store the offset values of the $(\text{id},\bm{d})$ and run length data blocks. The size of the offsets is defined as
\begin{equation}
    S_{offset} \triangleq h( \lceil \log_2{S_{pairs}} \rceil + \lceil \log_2{S_{RLE}} \rceil),
\end{equation}
where $h$ is the height of the image (i.e., number of rows), which is the number of block offsets we must store. The total compressed size of a RAGE encoded image, $I_C$, is
\begin{equation}\label{eq:gd_size_total}
    S \triangleq n_b l_b + S_{RLE} + S_{pairs} + S_{offset},
\end{equation}
where the first term, $n_b l_b$, is the size of $\mathcal{M}$. This expression excludes some parameters needed by the decoder to recreate the image. However, their size is negligible. Thus, Eq. \eqref{eq:gd_size_total} is used in Alg. \ref{alg:selection} to estimate the compressed size.

\section{Performance Evaluation}

\textbf{Comparison Schemes:} We compare performance of our developed methods to existing implementations of popular image compressors in C/C++. For lossless compression we consider four algorithms: 1) \textit{PNG} \cite{png}, using the single-file public domain STB library \cite{pnglib}; 2) \textit{JPEG-LS} \cite{jpegls}, using the Char-LS \cite{jpeglslib} implementation; 3) 
\textit{QOI} \cite{qoilib}, using the reference implementation of \textit{The Quite Okay Image format} \cite{qoi}; and 4) 
\textit{WebP} \cite{webp}, using the open source Libwebp library~\cite{webplib}.
For lossy compression, we compare to JPEG \cite{jpeg}. 

\textbf{Implementation:} RAGE methods are implemented Python 3. 
A practical RAGE decoder is developed in C++ to enable fast and efficient processing, to have fair decompression speed comparisons to the other algorithms, and for simple porting to embedded platforms. Reference implementations of each algorithm are compiled from source with gcc 9.4.0 using the same optimizations. 

\textbf{Test Platforms:} We evaluate important benchmarks on two different platforms.
As \textit{host platform} we use a Laptop with, 1.6 GHz i5 quad-core CPU. This is used to evaluate the compression performance and decoding speeds against the comparison algorithms. 
We use as \textit{target platform} a \textit{STM32U5G9J} \cite{stm32target}, 160 MHz 32-bit ARM Cortex-M33 processor to evaluate random-access capabilities.

\textbf{Data Sets:} The images data sets used for performance evaluations are a combination of commonly used data sets to evaluate image processing algorithms and images that could realistically by used in GUI applications. Images are of varying bit depths and of varying complexity to assess the compression performance on different classes of images. The used image data sets are summarized in Table \ref{dataset}.
Results presented are aggregated results of each data set.

\begin{table}
\small
\begin{center}
\begin{tabular}{|l|l|l|l|l|}
\hline
\textbf{Data set} & \textbf{Type} & \textbf{bpp} & \textbf{No. images} & \textbf{Ref.} \\
\hline
Icons* & Discrete & 24 / 32 & 923 & \cite{icons50, qoi_images} \\
\hline
Texture & Continuous & 24 & 22 &  \cite{qoi_images} \\
\hline
Alpha & Continuous & 32 & 190 & \cite{qoi_images} \\
\hline
Games* & Discrete & 24 & 619 & \cite{qoi_images} \\
\hline
Logos* & Discrete & 24 & 492 & \cite{logos} \\
\hline
CIFAR-10* & Continuous & 24 & 601 & \cite{learning} \\
\hline
Fonts* & Discrete & 24 & 703 & \cite{fonts} \\
\hline
Kodak & Continuous & 24 & 27 & \cite{kodak} \\
\hline
TGFX Stock & Discrete & 24 / 32 & 1523 & \cite{tgfx_stock} \\
\hline
\end{tabular}
\caption{Image data sets used. Data sets marked with * indicate a subset of the total referenced data set.}
\label{dataset}
\end{center}
\vspace*{-12pt}
\end{table}

\subsection{Lossless compression perfomance}

\textbf{Compression Ratio:}  Compression performance is measured using the \textit{compression ratio} defined as
\begin{equation}
    \text{CR} \triangleq \frac{\text{compressed size}}{\text{uncompressed size}}.
\end{equation}

\begin{figure}
    \centering
    \includegraphics[width=0.42\textwidth]{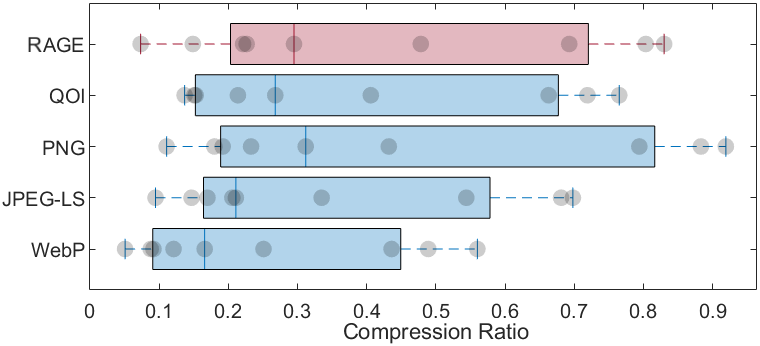}
    \caption{Box plots of lossless CRs for each image compressor.}
    \label{fig:box_plot}
    \vspace{-6pt}
\end{figure}

The distribution of the aggregated lossless CRs for all the data sets are summarized in Fig. \ref{fig:box_plot}, where the grey dots indicate CRs of individual data sets.
The median CR of GD is comparable to most popular standard image compression algorithms, except for WebP which outperforms all others.

\begin{figure}
    \centering
    \includegraphics[width=0.48\textwidth]{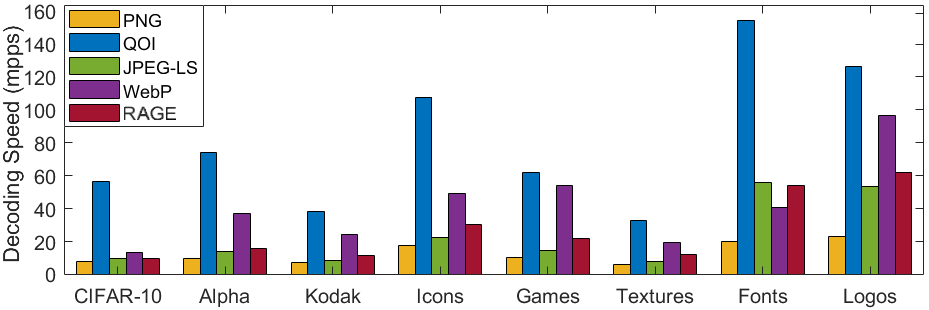}
    \caption{Decompression speed of lossless algorithms in megapixels per second (mpps).}
    \label{fig:decompression_speed_no_dr}
    \vspace{-6pt}
\end{figure}

\textbf{Decompression speed:} The aggregated decompression speeds on the host platform are summarized in Fig. \ref{fig:decompression_speed_no_dr}.
GD is consistently faster than the PNG implementation and often also outperforms JPEG-LS. For the discrete-tone data sets in particular (e.g., \textit{Fonts}, \textit{Logos}), a high throughout is achieved due to the effectiveness of the run-length encoding.

\begin{figure}
    \centering
    \rotatebox{90}{\quad \quad \quad \quad \, Distortion}
    \rotatebox{90}{\quad \quad\,\,\,\footnotesize{RAGE-Q}\quad\quad\quad\,\footnotesize{JPEG}}
    \begin{subfigure}[b]{0.22\textwidth}
        \centering
        \includegraphics[width=\textwidth]{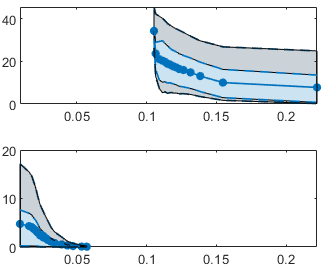}
        \captionsetup{justification=raggedright,singlelinecheck=false}
        \caption{\textit{Fonts}.\quad\quad CR}
        \label{fig:loss_fonts}
    \end{subfigure}
    \hfill
    \begin{subfigure}[b]{0.22\textwidth}
        \centering
        \includegraphics[width=\textwidth]{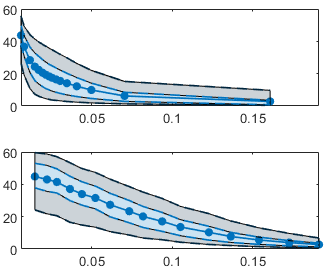}
        \captionsetup{justification=raggedright,singlelinecheck=false}
        \caption{\textit{Kodak}.\quad\,\,\,CR}
        \label{fig:loss_kodak}
    \end{subfigure}
    \caption{CR Vs. Distortion (MSE) of \textit{Fonts} and \textit{Kodak} data sets. Points on the solid line are the global distortion. Light blue region contains the 25th and 75th percentiles of local distortions. Grey region contains the 5th and 95th percentiles.} \vspace{-8pt}
    \label{fig:cr_vs_distortion}
\end{figure}

\subsection{Lossy compression performance}
Using the six datasets with 24 bpp (a restriction from JPEG), we evaluate the CR against the distortion of the reconstructed image measured by the \textit{mean square error} (MSE) defined as
\begin{equation}
    \text{MSE} \triangleq \frac{1}{n} \sum_{i=1}^{n}{(P_i-Q_i)^2},
\end{equation}
where $P_i$ denotes pixels of the original image and $Q_i$ the pixels of the reconstructed image. We evaluate the \textit{global distortion} of the image and the \textit{local distortion} by measuring the MSE of each $8 \times 8$ pixel block of the image to capture localized distortion regions.

We define the \textit{quality factor} for data sets where JPEG and RAGE-Q have overlapping compression performance as 
$\text{QF (CR)} \triangleq \text{MSE}_{RAGE-Q} \text{(CR)} / \text{MSE}_{JPEG} \text{(CR)}$
. RAGE-Q has a better global MSE for the same CR if QF(CR)$<$1.   

The aggregated CR and distortion of selected data sets is shown in Fig. \ref{fig:cr_vs_distortion}.
Fig. \ref{fig:loss_fonts} shows that the global and local distortion of RAGE-Q for a discrete-tone data set converges faster to zero than JPEG. In fact, zero distortion is achieved for a CR half the size of the smallest CR achievable by JPEG. For the continuous-tone data set in Fig. \ref{fig:loss_kodak} JPEG converges faster and achieves higher CRs at similar distortion levels.

\begin{figure}
    \centering
    \includegraphics[width=0.48\textwidth]{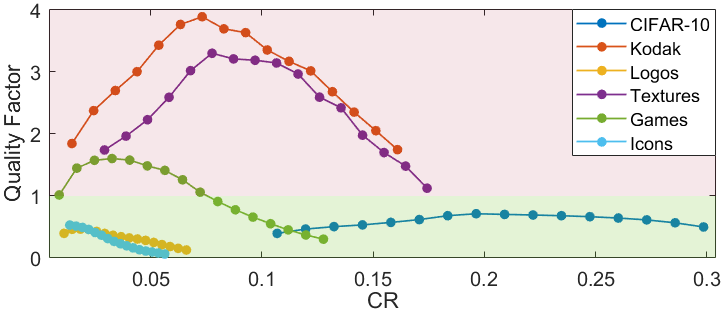}
    \caption{QF Vs. CR of data sets with overlapping compression performance. Points below 1 (the green zone) indicates lower distortion of RAGE-Q compared to JPEG at that CR.}
    \label{fig:quality_gain}
    \vspace{-12pt}
\end{figure}

Fig.~\ref{fig:quality_gain} shows RAGE-Q is 2 to 10+ times better in MSE for discrete-tone data sets, e.g., \textit{Icons}, \textit{Logos}.  JPEG performs better for most continuous-tone data sets, e.g., \textit{Textures}, \textit{Kodak}, except for \textit{CIFAR-10}, where RAGE-Q is strictly better. For the \textit{Games} data set, RAGE-Q is better at CR $>$ 0.07. The \textit{Fonts} data set is not depicted, but QF$=$0 given Fig. \ref{fig:loss_fonts} .

Table \ref{visual_example} shows a visual comparison, where RAGE-Q outperforms JPEG in both CR and MSE. JPEG's smearing and artifacts are more visible and evident than the color quantization introduced by RAGE-Q even at similar MSEs.


\begin{table}
\begin{center}
\setlength\tabcolsep{0.1pt}
\begin{tabular}{p{1.5cm}>{\centering}p{1.7cm}>{\centering}p{1.7cm}>{\centering}p{1.7cm}>{\centering\arraybackslash}p{1.7cm}}
\textbf{JPEG}: &
\includegraphics[width=\linewidth,valign=m]{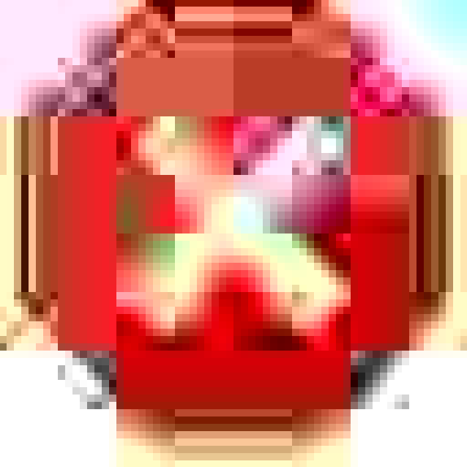} & \includegraphics[width=\linewidth,valign=m]{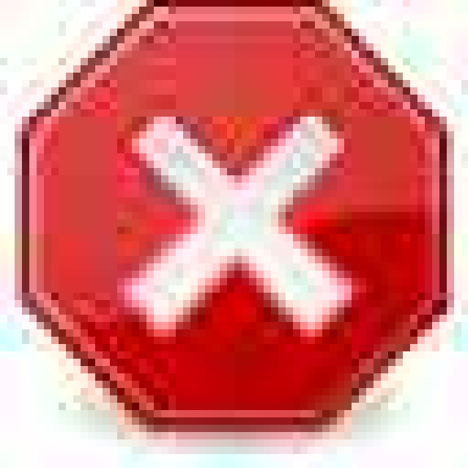} & \includegraphics[width=\linewidth,valign=m]{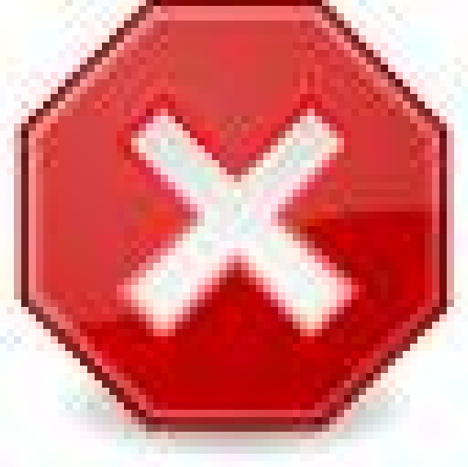} & \includegraphics[width=\linewidth,valign=m]{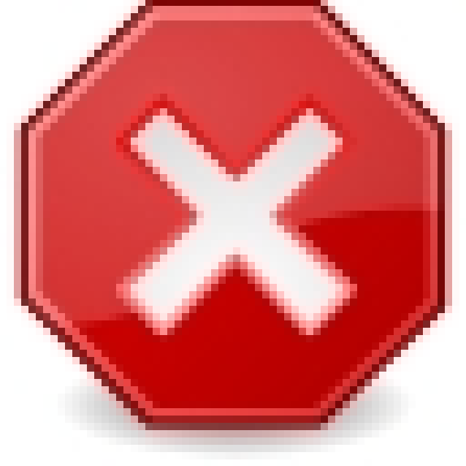}\\
CR: & 0.065 & 0.105& 0.149 & 0.447 \\
MSE: & 80.4 & 60.8 & 47.3 & 22.4 \\
\textbf{RAGE-Q}: & 
\includegraphics[width=\linewidth,valign=m]{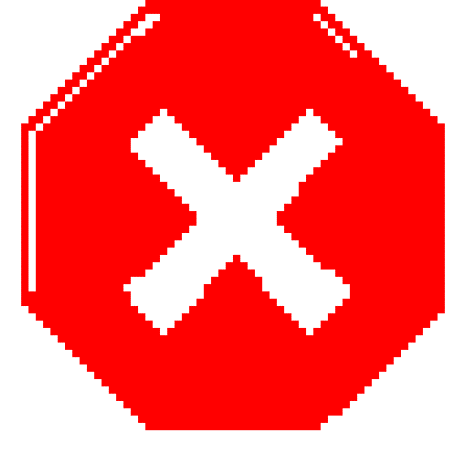} & \includegraphics[width=\linewidth,valign=m]{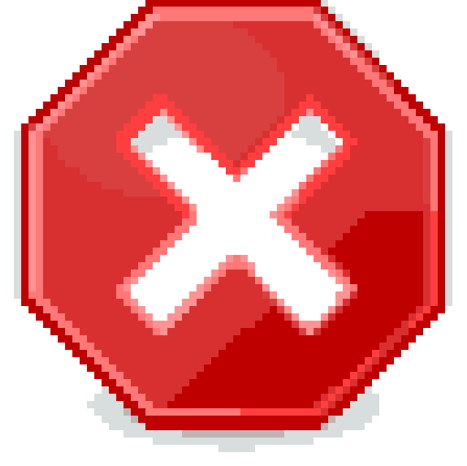} & \includegraphics[width=\linewidth,valign=m]{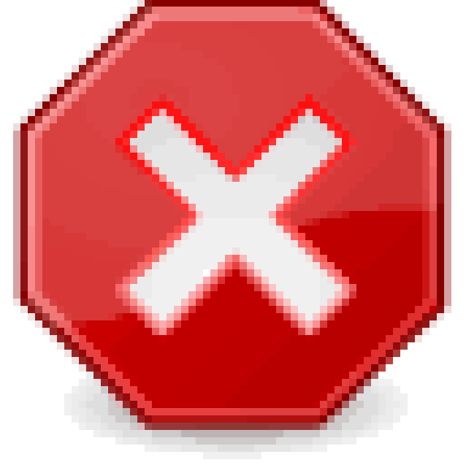} & \includegraphics[width=\linewidth,valign=m]{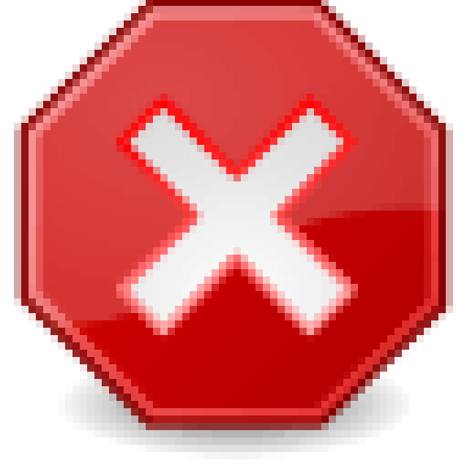}\\
CR: & 0.029 & 0.073 & 0.135 & 0.245 \\
MSE: & 72.86 & 50.85 & 11.19 & 1.33 \\
\end{tabular}
\end{center}
\vspace{-16pt}
\caption{Visual comparison of JPEG and RAGE-Q}\label{visual_example}
\end{table}

\subsection{Random Access Performance}
The random access queries considered are rectangular sub-regions of an image, described by the tuple \texttt{<x, y, w, h>}, where $(x,y)$ is the coordinate of the top-left corner and $w$ is the width and $h$ is the height. These are typical in embedded GUIs, e.g., TouchGFX~\cite{tgfx_stock}.
By design, our scheme supports random-access queries without decompression, thus making them less costly. The total time it takes to seek through the compressed data to locate the $(\text{id},\bm{d})$-pairs of a query area is denoted $t_{seek}$. We define the average (AVG) $t_{seek}$ as the aggregated time it takes to seek past a chunk (i.e., pixel). This has been measured by simulating queries of increasing sizes starting by the left-most column of pixels and expanding until covering the whole image. Table \ref{seek_measurements} shows the AVG $t_{seek}$ and \textit{decoding time per pixel} (dtpp) of a subset of 16 image per data set on the target platform.
\begin{table}
\small
\begin{center}
\begin{tabular}{|l|p{2.5cm}|p{3.2cm}|}
\hline
\textbf{Data set} & Average Seek Time \textbf{$\bm{t_{seek}}$ [ns]} & Decoding Time Per Pixel \textbf{dtpp [ns]} \\
\hline
Fonts & 15.6 & 274\\
\hline
Logos & 12.3 & 277\\
\hline
Icons & 25.6 & 277\\
\hline
Games & 21.6 & 619\\
\hline
Alpha & 10.0 & 834\\
\hline
Textures & 9.9 & 1049\\
\hline
Kodak & 13.0 & 1123\\
\hline
CIFAR-10 & 40.6 & 1226\\
\hline
\end{tabular}
\caption{Average $t_{seek}$ and dtpp measured on target platform}
\label{seek_measurements}
\end{center}
\vspace*{-24pt}
\end{table}
As the AVG $t_{seek}$ is much smaller than the dtpp, the seeking of compressed data to locate any random query area will be much faster than the actual decoding of chunks to recreate it. 

\section{Conclusions}
We introduced RAGE as a flexible image compressor for fast, pixel-level random access without the need for decompression of the entire image, which is suitable for resource constrained devices. 
Although RAGE is studied in the context of GUI, its impact can be wide-ranging. 
Future work will study its potential for image processing tasks, e.g., machine learning tasks, to be executed in an embedded device. Although tinyML support and accelerators are becoming commonplace, storage and fast data access will be a critical bottleneck. Future work will also consider expansions to support animated graphics in embedded devices, e.g., a GIF \cite{gif} alternative.

\bibliographystyle{IEEEbib}
\bibliography{main}

\end{document}